\documentclass[prb,twocolumn,showpacs,preprintnumbers,amsmath,amssymb]{revtex4}
\usepackage{graphicx}
\usepackage{dcolumn}
\usepackage{bm}
\usepackage{slashed}
\newcommand {\sll}[1]{{\slashed #1}}

\usepackage{trfsigns}
\newcommand \be{\begin{eqnarray}}
\newcommand \ee{\end{eqnarray}}
\newcommand \ba{\begin{align}}

\newcommand {\V}[1]{{\bf #1}}

\newcommand {\ov}[1]{\overline{#1}}

\DeclareMathOperator{\Tr}{Tr}

\begin{document}
\title{Exploring anomalies by many-body correlations}
\author{K. Morawetz$^{1,2}$
}
\affiliation{$^1$M\"unster University of Applied Sciences,
Stegerwaldstrasse 39, 48565 Steinfurt, Germany}
\affiliation{$^2$International Institute of Physics- UFRN,
Campus Universit\'ario Lagoa nova,
59078-970 Natal, Brazil}
\begin{abstract}
The quantum anomaly can be written alternatively into a form violating conservation laws or as non-gauge invariant currents seen explicitly on the example of chiral anomaly. By reinterpreting the many-body averaging, the connection to Pauli-Villars regularization is established which gives the anomalous term a new interpretation as arising from quantum fluctuations by many-body correlations at short distances. This is exemplified by using an effective many-body quantum potential which realizes quantum Slater sums by classical calculations. It is shown that these quantum potentials avoid the quantum anomaly but approaches the same anomalous result by many-body correlations. A measure for the quality of quantum potentials is suggested to describe these quantum fluctuations in the mean energy. Consequently quantum anomalies might be a short-cut way of single-particle field theory to account for many-body effects. This conjecture is also supported since the chiral anomaly can be derived by a completely conserving quantum kinetic theory. 
\end{abstract}

\maketitle

\section{Introduction}

\subsection{Anomalies}
Anomalies are a puzzling discovery in quantum field theory. Certain classical symmetries and conservation laws are broken if the fields become quantized and have been named anomalies. They have a long history starting with investigations of pion decays \cite{St49,Ad69,BJ69}, for an overview see \cite{Bert00}. These anomalies are important for the description of a variety of experiments. Besides the neutral pion decay also the spontaneously broken axial $U(1)$ symmetry in QCD should be mentioned seen in no parity doubling of baryons and no related Goldstone boson \cite{tH76} as well as the Kaon decay \cite{WZ71}. The chiral anomaly as breaking of chiral symmetry \cite{Ad69,LEUT94} has recently gained a renewed interest in condensed matter physics as excitation of chiral mass-less Fermions in the class of Weyl semi-metals \cite{Xu613,Lu622,Lv15,Hu15,We15}. It was predicted in \cite{Wa11} and experimentally interpreted \cite{Hu15,Xu17,gooth17} as having observed chiral anomaly. This has led to an enormous theoretical activity \cite{BKY14,JJM19,FB00} describing e.g. anomalous transport \cite{ACF98,Ki13,La16,GMSS18,M16,M18}, the relation of chiral anomaly and quantized Hall effects \cite{Abo85,MHR15} up to chiral heat effect \cite{KN12}.

The anomalies lead to anomalous Ward identities \cite{Bard69,DJ74} and destroy the gauge invariance \cite{WZ71,PRE91} seen also in trace anomalies \cite{BN93}. One can either formulate the theory consistent or covariant \cite{BZ84,RDS97,AD98,EK00,TE17} dependent whether one accepts alternatively violation of gauge invariance or violation of conservation laws. Furthermore the acceptance of anomalies does not guarantee renormalizability of the theory \cite{GJ72,BRS81}. 

Therefore it is highly desirable to formulate the theory free of anomalies or re-describe the experimental facts by a consistent theory. In this respect various anomaly cancellations have been proposed. In electro-weak interaction of the standard model the demand of anomaly-free formulation restricts the fermionic content \cite{Bert00}. Nonlocal counter-terms of gauge fields have been used to compensate anomalies \cite{Kr85}. Extending the initial phase space \cite{MOS89} or using higher dimensions, cancellations \cite{GSW85} have been also worked out. 
Due to the axial non-conservation sometimes the chiral anomaly is called also mixed axial-gravitational anomaly and claimed to violate Lorentz symmetry \cite{ZB12,JHK17,Xu17,gooth17}.  In \cite{DelCima2017} it has been shown that a proper subtraction scheme of the infrared divergences shows that corresponding extra terms do not appear. In fact the Lorentz-invariant chiral kinetic theory can be derived from the quantum kinetic approach \cite{GLPWW12,CPWW13,MT14,GPW17,HPY17,HSJLZ18,M18} leading to the chiral anomaly by many-body correlations. A hint that the anomalies can be possibly explained by  many-body effects is also the observation that they can be described by the Dirac sea \cite{NN83,VL90,ABH93}.

We conclude that besides the well worked out mathematical appearances of anomalies as triangle graphs in field theory\cite{Bard69,Ad69,BZ84,BN93,Bert00,Fuj04},  the physical origin of anomalies is still a matter of debate. Therefore it is the motivation of this paper to draw the attention to a possible different scenario. Given the experimental facts, we believe that the anomalies describe real physics. However, it is the question whether it has to appear as anomalous or whether the same physics can be described by ordinary means. Let us employ an analogy. Pairing in superconductors is conveniently described by anomalous propagators to achieve the Gorkov equations or correspondingly the Beliaev equations for Bose condensates. These propagators violate the number conservation and are inconsistent in the above sense. Nevertheless they lead to correct equations. Recently it has been shown that one can arrive at the same equations by a completely conserving theory of multiple corrected T-matrix \cite{L08,Mo10,SLMMM11} with equivalent results \cite{M10,MML13}. This illustrates that the anomalous propagators are a theoretical short-cut to the right result though adopting inconsistent steps. 

Analogously we will propose here that the quantum anomaly might be a short cut way to describe correct physics. We will propose to consider it as a many-body correlation phenomenon. Let us illustrate this in terms of the well-discussed chiral anomaly before we restrict to the non-relativistic case in the paper. 

\subsection{Field theoretical approach}
Relativistic Fermions with zero mass and consequently linear dispersion have a definite chirality by parallel or anti-parallel spin and motion directions \cite{W29}. The left and right-handed projections are realized by $(1\mp\gamma_5)/2$ with $\gamma_5=i\gamma^0\gamma^1\gamma^2\gamma^3$. The chiral or axial transformation
\be 
\Psi'(x)={\rm e}^{i\alpha(x) \gamma_5}\Psi(x)
\ee
leads to the axial current $J_5=\ov {\Psi} \gamma^\mu \gamma^5\Psi$ which changes the classical action $S'=S+\int \alpha(x) \nabla_\mu J_5^\mu$. This results into the conservation law
\be
\nabla_\mu J_5^\mu=2i m \ov \Psi \gamma^5\Psi\to 0,\quad {\rm for}\quad m\to 0
\ee
for mass-less Dirac particles.
The quantum averaging in contrast,
\be
\langle \partial_\mu J_5^\mu\rangle&=& 2i m \langle \ov \Psi \gamma^5\Psi\rangle \to  {e^2\over 16 \pi^2\hbar^2 c}\varepsilon^{\mu\nu\alpha\beta}F_{\mu\nu}F_{\alpha\beta}
\nonumber\\
&=&{e^2\over 2 \pi^2\hbar^2} {\V E \cdot \V B}\quad {\rm for}\quad m\to 0,
\label{anom}
\ee
shows a non-vanishing anomalous term obviously due to quantum fluctuations in the average.

The origin is best seen from Pauli-Villars regularization where we subtract from the Dirac Lagrangian for $\Psi$ a massive (${ M\to\infty}$) field $\Phi$ \cite{kaf91,Bert00,Fuj04} 
\be{\cal L}=i{\bar \Psi} 
(\sll  \partial -ie\sll A) \Psi-i{\bar \Phi} (\sll \partial -ie \sll A) \Phi +{ M} {\bar \Phi} \Phi.
\ee
For the chiral current we calculate ${\rm Tr} \gamma^5 G_{12}$ with $(i\sll \partial_1-{ M}+e\sll A_1)G_{12}=-\delta_{12}$ iteratively by $G=G_0+G_0 e\sll A G$ and $G_0=-(\sll p-{ M})^{-1}$. Due to the trace the first non-vanishing terms are of fourth-order
\be
&&\langle \partial_\mu J_5^\mu\rangle=8 \epsilon^{\kappa \lambda \mu \nu}\int {d^4qd^4 r \over (2\pi)^4} {\rm e}^{irx} e r^\nu A_{q-r}^\mu q^\lambda  eA_{-q}^\kappa
\nonumber\\
&&\times \int {d^4p\over (2\pi)^4}{{ M}^2\over (p^2-{ M}^2)((p+q)^2-{ M}^2)((r+p)^2-{ M}^2)}
\nonumber\\
&&=-{1\over 4 \pi^2\hbar^2} \epsilon^{\kappa \lambda \mu \nu}\partial_x^\nu eA^\mu\partial_x^\lambda e A^\kappa
\nonumber\\
&&=-{e^2\over 16 \pi^2\hbar^2} \epsilon^{\kappa \lambda \mu \nu}F^{\nu\mu}F^{\lambda \kappa}={e^2\over 2 \pi^2\hbar^2} { \V E \cdot \V B}
\label{rate}
\ee
where one calculates the integral in the $M\to\infty$ limit with $\int d^4p /(p^2+M^2)^3=\pi^2/2M^2$. This means it comes from divergences up to fourth adiabatic order (renormalization) which can be expressed by anomalous triangle graphs \cite{J12,BKY14,L16} known as Adler-Jackiw-Bell anomaly \cite{Ad69,BJ69,NN83}.   The origin is clearly the behaviour at small distances or large momenta. 
This chiral anomaly can be based on anomalous Ward identities which quantum vector or axial vector fields obey. Only exclusively one of them can be made normal \cite{Ba69}. In fact, the rate of chirality (\ref{rate}) can be rewritten explicitly \cite{M19}
\be
\partial_t n_5+{\V \nabla}\cdot (\V j+\V j_{\rm anom})=0.
\label{3}
\ee
Using the vector and scalar potentials $\V B=\nabla\times \V A, \V E=-\dot {\V A}-\nabla \phi$, the anomalous current 
\be
\V j_{\rm anom}={e^2\over 2 \pi^2\hbar^2}\left (\frac 1 2 \dot {\V A}\times \V A-\phi {\V \nabla}\times \V A \right )
\label{j}
\ee
is non gauge-invariant. So one can choose either to accept a non-conserving rate equation (\ref{rate}) or alternatively a conserving rate equation (\ref{3}) but with a non gauge-invariant current (\ref{j}).

\subsection{Many-body approach}

The same anomalous result (\ref{anom}) can be obtained by many-body effects without anomalous behaviour. Heuristically it can be seen easily \cite{Fuk08} considering a parallel electric and magnetic field which changes the chirality. The Fermi momentum of the right-handed Fermions increases in the electric field
$
p_{\rm F}=eEt
$
with opposite direction for left-handed ones. The density of left and right-handed Fermions is the product of the longitudinal phase-space density, $dN_R/dz=p_{\rm F}/2\pi\hbar$, and the density of Landau levels in traverse direction, $d^2N_R/dx dy=eB/2\pi\hbar$, such that the rate of chirality $N_5=N_R-N_L$ is
\be
{d n_5\over d t}={d^4 N_5\over d t d^3x}=2 {\dot p_{\rm F}\over 2 \pi \hbar}{e B\over 2 \pi \hbar}={e^2\over 2 \pi^2\hbar^2} \V E \cdot \V B
\label{rate1}
\ee
which agrees with (\ref{anom}).
We see that a completely conventional reasoning leads to the same result as obtained by anomaly. The quantum kinetic derivation of this result without any non-conserving assumptions can be found in \cite{M18}. 

The $\V E\V B$ term is also the basis of the experimental interpretation \cite{Hu15,Xu17,gooth17} of having observed chiral anomaly and breaking of conservation laws like mixed axial-gravitational anomaly. The electrodynamics assuming explicitly a chiral breaking term has been treated in \cite{QCH17}. The well investigated path from symmetry-violating assumptions to final `non-conservation` form  \cite{ZB12a} was in a sense misleading if one sees it as an unique signal of violation of conservation laws. One cannot conclude backwards from the observed term (\ref{anom}) to a symmetry-breaking field-theoretical assumption since (\ref{rate}) appears also by a conserving theory without the described field-theoretical assumptions \cite{M18}. 

\subsection{Conjecture and outline}

If the physical origin of the anomalous term is the behaviour at small distances and the quantum fluctuations by many-body correlations, one should be able to get the anomalous results by ordinary many-body treatments. We want here to investigate more in detail how the quantum averaging over the wave function can consistently be performed within a many-body treatment. In fact we will show that a proper treatment of such many-body averaging renormalizes the divergence at small distances and no anomaly appears. The term (\ref{3}), however, shows up nevertheless since it has a many-body origin. Therefore the conjecture is proposed that the anomalies are short-cut ways of single-particle field theory to a many-body effect. In the single-particle treatment they appear as anomalous, in many-body treatment they appear naturally without anomaly.

The outline of the paper is as follows. In the next chapter II the nonrelativistic anomalies are re-derived by many-body correlations and the statistical sum. Conditions are discussed dependent on dimensionality, power law of potentials and perturbation order. This results into an anomalous energy shift of $\frac 1 8 E_0$. In chapter III we show that the use of quantum potentials avoid this anomaly but leads naturally to this energy shift due to the finite value of the potential at small distances. This finite value is caused by quantum fluctuations which are represented by quantum potentials discussed in binary and ternary order. Chapter IV summarizes the results. Appendix A provides integrals occurring in the treatment of chapter II. Appendix B discusses the derivation of quantum potentials on binary and ternary level separately for Maxwellian and for Fermi correlations.

\section{Anomaly by statistical sum\label{eanom}}

First we observe that the averaging over a many-body statistical operator with kinetic $\hat E$ and potential $\hat V$ energy can be written as inverse La-Place transform
\ba
&z={\rm Tr}\, {\rm e}^{-\beta (\hat E+\hat V-\mu \hat N)}={\rm Tr} \!\!\!\!\!\int\limits_{-i \infty+\epsilon}^{i\infty +\epsilon}\!\!{dM\over 2\pi i}  {\rm e}^{\beta (M\!+\!\mu \hat N)} {1\over \hat E +\hat V+M}
\nonumber\\
&=
\!{\rm Tr}\!\!\! \!\!\int\limits_{-i \infty+\epsilon}^{i\infty +\epsilon}\!\!\!{dM\over 2\pi i} {\rm e}^{\beta (M\!+\!\mu \hat N)}\!\left ({1\over \hat E\!+\!M}-{1\over \hat E\!+\!M}V{1\over \hat E\!+\!M}\pm....\right )
\label{laplace}
\end{align}
where we expand with respect to the interaction and assume that it commutes with the number operator.
The vanishing statistical averaging corresponds to the high-temperature limit $\beta\to 0$ or alternatively to $M\to \infty$. The latter one takes the role of the large mass of Pauli-Villars regularization. To see this, consider the pure quantum state expectation \cite{ks96} in $D$-dimensions
\ba
{\rm Tr^{\rm QM} \hat A}\!=\!\!\int \!\!\!{d^D p\over (\!2\pi \hbar\!)^D} \langle p|\hat A|p \rangle=\!\!\int\!\!\! {d^D \!x d^D\! p\over (2\pi \hbar)^D}{\rm e}^{-{i\over \hbar} \V p \V x}\langle x| \hat A|x'\rangle {\rm e}^{{i\over \hbar} \V p \V x'}. 
\label{QM}
\end{align}
If the observable $\hat A$ does not contain any explicit $\hbar$, one expects
$(2\pi \hbar)^D {\rm Tr \hat A}-\lim\limits_{\hbar\to 0}(2\pi \hbar)^D {\rm Tr \hat A}=0$. A violation of this zero represents the quantum anomaly. 

To investigate it further, we combine the many-body mixed state (\ref{laplace}) and the quantum-mechanical pure state (\ref{QM}) averaging
\ba
W^{(1)}&=(2\pi \hbar)^Dz-\lim\limits_{\hbar\to 0} (2\pi \hbar)^D z
\nonumber\\&
=\sum\limits_{n=0}^\infty(-1)^n\int {d M\over 2\pi i}{\rm e}^{\beta M} W_n
\label{10}
\end{align}
with 
\ba
W_n&=\sum\limits_{n_1}n_1^n{\rm e}^{\beta \mu n_1}\!\int \!\! d^Dp d^Dx
\nonumber\\
&\times \left [{1\over n_1{(\V p\!-\!i\hbar \partial_{\V x})^2\over 2 m}\!+\!M}V(x){1\over n_1{(\V p\!-\!i\hbar \partial_{\V x})^2\over 2 m}\!+\!M}V... \right .
\nonumber\\
&-\, \left . {1\over n_1{(p)^2\over 2 m}+M}V(x){1\over n_1{(p)^2\over 2 m}+M}V\right ]
\label{Wn1}
\end{align}
where we have used the non-relativistic kinetic energy $\hat E=\hat p^2/2m$ as example. The relativistic dispersion works analogously. We have commuted the $\exp{(-i\V p\V x/\hbar)}$ factor in (\ref{QM}) through the kinetic energy factors. The sum runs over the number of Fermions $n_1=0,1$ or Bosons $n_1=0,1,2...$.

\begin{figure}
\includegraphics[width=7cm]{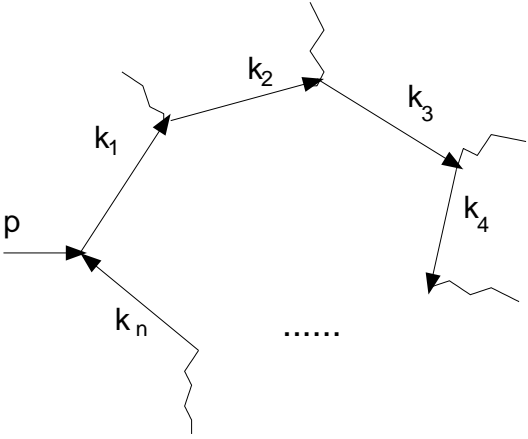}
\caption{\label{ringgraph}Propagators interacting with external potentials in a ring according to (\ref{In}).}
\end{figure}

Now we assume a momentum dependence of the Fourier transformed potential in the form of a power law 
\be
V(q)=E_0 a_0^D c_d q^{-{ \alpha}}
\ee
with a typical energy $E_0$ and length scale $a_0$. This would be the Bohr radius $a_0=a_B$ and $E_0=Ryd =e^2/4 \pi \epsilon_0a_B$ for Coulomb potentials in D=1,2,3 dimensions with the numerical factor $c_d=2^{D-1}\pi $. The Fourier transformation (\ref{Wn1}) in dimensionless momentum $\V k\to \V k\sqrt{n_1/2 m M}$ reads then
\be
W_n={(2 \pi \hbar)^D c_d^n\over 2^{\frac n 2 (\alpha-D)}}
E_0^{s+1}\sum\limits_{n_1}
{{\rm e}^{\beta \mu n_1}\over n_1^{s+1}}M^s \,
 I_n^{(D)}(\alpha)
\ee
with
\be
s=\frac n 2 (D-\alpha-2)-1
\ee
and
\ba
&I_n^{(D)}(\alpha)=\!\!\int\!\!\! {d^Dk_1\over (2\pi)^{D}k_1^\alpha}...{d^Dk_n\over (2\pi)^{D}k_n^\alpha}{d^Dp\over 1 \!+\!p^2} \delta\left (\Sigma_{i=1}^n \V k_i\right )
\nonumber\\
&\times\!\!\left [
{1\over 1\!+\!(\V p\!+\!\V k_1)^2}...{1\over 1\!+\!(\V p\!+\!\V k_1\!+\!...\!+\!\V k_n)^2}\!-\!{1\over (1\!+\!p^2)^n}
\right ].
\label{In}
\end{align}
One can understand this integral as a ring-graph of $n$ propagators interacting $n$ times with external potentials
as illustrated in figure~\ref{ringgraph}, first considered in \cite{Bard69}.

The expression $W^{(1)}$ describes the anomaly in the normalization of the statistical operator or the completeness. The corresponding term for the mean energy is conveniently expressed as $W^{(H)}=-\partial_\beta W^{(1)}$. If we are interested in the anomaly of the momentum, we have an additional momentum factor which requires $s \to s+1/2$ and a fore-factor $\sqrt{2}\hbar/a_B$ as well as modified integrals (\ref{In}) by an additional $p$ factor.

Performing the inverse La Place transform (\ref{10}) we obtain finally
\be
W^{(1)}=\sum\limits_{n_1,n=1}^\infty {(-c_d)^n(2 \pi \hbar)^D\over 2^{\frac n 2 (\alpha-D)}}{{\rm e}^{\beta \mu n_1} I_n^{(D)}(\alpha)\over \Gamma(-s)(E_0 \beta n_1) ^{1+s}}.
\label{W}
\ee

Let us now discuss the  $M\to \infty$ or ${ \beta} \to 0$ analogous to the Pauli-Villars regularization. This will produce a non-vanishing anomaly only for certain combinations of dimensions $D$, power of the momentum in the potential $\alpha$ and the order of terms $n$ in the sum (\ref{W}). 

One sees the dependence on $\beta$ as
\be
W^{(1)}=(2\pi\hbar)^D \delta z&\sim& \beta^{-1-s}\nonumber\\
W^{(H)}=(2\pi\hbar)^D \delta H&\sim& (1+s)\beta^{-2-s}\nonumber\\
W^{(p)}=(2\pi\hbar)^D \delta p&\sim& \beta^{-\frac 3 2-s}
\ee
which means that one gets a nonzero anomaly for the normalization $\delta z$, the energy $\delta E$ and the momentum $\delta p$ only for the combinations 
\be 
\delta z&\ne& 0: \forall n, D=2+\alpha \nonumber\\
\delta H&\ne& 0: n=1, D=\alpha \, \,{\rm or}\,\, n=2, D=\alpha+1 \nonumber\\
\delta p&\ne& 0: n=1, D=2+\alpha.
\label{cases}
\ee
For any sensible potential the relation for the anomaly in the normalization $\delta z$  is not fulfilled. Also the case of momentum anomaly is zero due to the vanishing integral (\ref{I1Da}). We get therefore the anomaly only for the mean value of the energy
\ba
\langle H\rangle&={\Tr{\hat H\hat \rho}\over z}\!+\!\Delta H, \,\,{\rm for}\,\, n=1, D=\alpha\, {\rm or}\, n=2, D=\alpha\!+\!1.
\end{align}
Let us discuss this anomaly in each dimension.
In three dimensions
 ${ D} =3$ and Coulomb ${ \alpha}=2$ interaction, the integral (\ref{In}) is given by (\ref{I322}) and one obtains
\be
\Delta H={W^{(H)}\over (2\pi\hbar)^3}=\frac 1 8 E_0
\label{deltaE}
\ee
a result reported in \cite{ks96} with a factor of $2$ due to spin which we omit here. Summarizing, the mean energy $H$ shows a quantum anomaly but not the normalization $z$. 

In two dimensions with a potential of ${ \alpha}=1$ one finds analogously 
\be
W_n^{(1)}\sim{ \beta}^{n/2}\to 0\,{\rm and}\, W^{(H)}_n \sim{ \beta}^{n/2-1}\ne 0\,{\rm  for}\, n=2
\ee
and for ${ \alpha}=2$
\be
W_n^{(1)}\sim{ \beta}^n\to 0\, {\rm and}\, W^{(H)}\sim { \beta}^{n-1} \ne 0\,{\rm for}\, n=1.
\ee

In one dimensions we do not have any anomaly for any potential. Indeed, considering  $\alpha=1$
we have
\be 
W_n\sim{ \beta}^n\to 0, {\rm and}\, W^{(H)}\sim { \beta}^{n-1} \ne 0,{\rm for}\, n=1 
\ee
but $I_1^{(1)}(1)=0$ due to (\ref{i1da}).
For $\alpha=2$ all anomalies vanish in 1D due to
$W_n\sim { \beta}^{\frac 3 2 n}\to 0$ and $W^{(H)}_n\sim { \beta}^{\frac 3 2 n-1} \to 0$.
The Fourier transform of the Coulomb potential in 1D is $\alpha=0$ but $V(q) \sim {\rm sign}(q)$ which leads to $I_2^{(1)}(0)=0$ as well.  

We can conclude that in all discussed cases the anomalies appear due to the large momentum (short distance) divergence of the potential. The results depend on dimensions and the anomaly appears in 2D and 3D but not in 1D and can be understood as a certain limit of the statistical sum. In the following we restrict to the discussion of three dimensions.

\section{Effective quantum potential}

We can now state that a finite behaviour of the potential at small distances realized by
quantum potentials cures such anomalies since we will see that they lead to $\alpha=4$ which avoids all cases of anomalies (\ref{cases}). These effective quantum potentials appear if we represent the many-body binary quantum correlations by a classical calculation with an effective quantum potential. Formally there are two ways to achieve this goal. The first way constructs the quantum potential such that the two-particle quantum Slater sum is correctly represented by the 
classical one with the help of the quantum potential. This results into the Kelbg 
potential for Maxwellian Coulomb systems \cite{K64,K641,EHK67,KK68}
\be
 V_{2}(r)&=&{E_0 a_0\over r}
\left [
1-{\rm e}^{-\left ({r\over l}\right )^2}+\sqrt{\pi}\,{r\over l}\, {\rm erfc}
\left ({r\over l}\right ) 
\right ]
\nonumber\\
&=&E_0 a_0
\left \{
\begin{array}{c}
{1 \over r}+o(r^{-3})\cr
{\sqrt{\pi}\over l}-{r\over l^2}+o(r^2)
\end{array}
\right .\label{Kelbg}
\ee
where the coordinate is scaled by the thermal wave length $l^2=2 \hbar^2/m T$. Improvements and systematic applications 
can be found in \cite{kker86,KK681,OVE00}.

The second way is to use a statistical equivalence of quantum $N$-particle systems with an $N+1$-particle classical system \cite{Mq01}. There it was found that the quantum potentials are just successive convolutions of the (Coulomb) potential $V^c(x)$ with the binary distribution $\rho^{(2)}(x)$. If we use the non-degenerate Maxwell correlation
\be
\rho^{(2)}(r)\!&=&\!\!\int \!\!\!{d p\over (2 \pi \hbar)^3} {\rm e}^{i p r/\hbar} {\rm
  e}^{-\beta {p^2\over 2 m T}}={1\over \pi^{3/2} l^3}{\rm e}^{-r^2/l^2}
\label{rho1}
\ee
the Kelbg potential (\ref{Kelbg}) appears as 
\be
 V_{2ab}(r)&\propto& \sum\limits_{c}\int {d x_1} \rho^{(2)}_{bc}(x_1) V^c_{cb}(x_1) V^c_{ca}(x_1+r)
\ee
with the quantum number, e.q. being charges, indicated by latin subscripts.
As calculated in appendix~\ref{eff}, using the Fermi function at $T=0$, we obtain instead of (\ref{rho1}) the potential
\be
V_{2\rm f}(r)\!&=&\!
{E_0 a_0\over 2 r}\!\left [\!
2\!+\!{\pi r\over 2 l_{\rm F}}\!-\!{\cos{r\over l_{\rm F}}}\!-\!{l_{\rm F}\over r} \sin{r\over l_{\rm F}} \!-\!{r\over l_{\rm F}} Si\left (\!{r\over l_{\rm F}}\!\right )
\!\right ]
\nonumber\\
&=&E_0 a_0
\left \{
\begin{array}{c}
{1 \over r}+o(r^{-3})\cr
{\pi\over 4 l_{\rm F}}-{r\over 6 l_{\rm F}^2}+o(r^2)
\end{array}
\right .
\label{Fermi}
\ee
where the coordinate scales with the inverse Fermi momentum $l_{\rm F}=\hbar/p_{\rm F}$ and the sinus integral is $Si(x)=\int_0^xdt \sin{t}/t$.

This scheme allows to construct besides the binary quantum potential also the next (ternary) order 
\ba
& V_{3a b}(r)\propto
\sum\limits_{cd}\int {d x_1 dx_2} \rho_{dc}(x_1) V^c_{cd}(x_1)
V^c_{cb}(x_1+x_2) \nonumber\\
&\times
\rho_{bd}(x_2) V^c_{ca}(x_1+x_2+r)
\label{ternary}
\end{align}
which leads with (\ref{rho1}) to
\ba
V_{3}(r)&=
{E_0 a_0\over r}
\left [
{\rm erf}^2\!\left (\!{x\over
      \sqrt{2}}\!\right )\!+\!
  {2^{3/2} x\over \sqrt{\pi}} \!\int\limits_{ x}^{\infty} \!{dz \over z} {\rm
      e}^{-{z^2\over 2}} {\rm erf}\!\left ({z\over \sqrt{2}} \!\right )\right ]
\nonumber\\
&=E_0 a_0
\left \{
\begin{array}{c}
{1 \over r}+o(r^{0})\cr
{\sqrt{8}\,{\rm ln}(1+\sqrt{2})\over \sqrt{\pi} l}-{2r\over \pi l^2}+o(r^2)
\end{array}
\right .
\label{tern}
\end{align}
calculated in appendix~\ref{eff}. The corresponding ternary order for Fermi correlations reads
\ba
V_{3\rm f}(r)&=
{E_0 a_0\over r}{4\over \pi^2+4}\int\limits_0^1 {d x\over x} 
\left [x+\left (1 -x^2\right ) {\rm artanh} x\right ]
\nonumber\\
&\times\left [
2+\pi {x r/ l_{\rm F}}-2 \cos{({x r/ l_{\rm F}})}-2x {r\over l_{\rm F}} {\rm Si}\left ({x r/ l_{\rm F}}\right )
\right ]
\nonumber\\
&=E_0 a_0
\left \{
\begin{array}{c}
{1 \over r}+o(r^{0})\cr
{4\pi \,(1+2 {\rm ln} 2)\over 3(4+\pi^2)l_{\rm F}}-{2 r\over (4+\pi^2)l_{\rm F}^2}+o(r^2)
\end{array}
\right ..
\label{ternf}
\end{align}
  
\begin{figure}
\includegraphics[width=8cm]{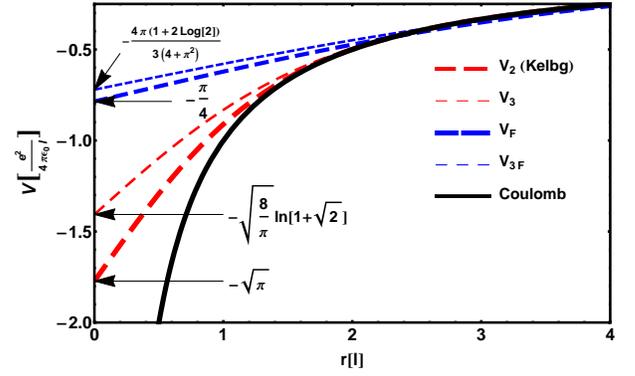}
\caption{
\label{comp} The comparison of the Kelbg potential (\ref{Kelbg}) of binary correlations together with the next (ternary) order correlation potential (\ref{tern}), the Fermi potential (\ref{Fermi}) and the Coulomb potential. For the Kelbg and ternary potential the scale is $l=\hbar\sqrt{2\beta/m}$ and for the Fermi potential $l=l_{\rm F}=\hbar/p_{\rm F}$. The finite value at zero distance is explicitly indicated.
}
\end{figure}

The results (\ref{Fermi}) and (\ref{ternf}) are not yet reported while the ones (\ref{Kelbg}) and (\ref{tern}) had been presented in \cite{Mq01}.
The quantum potential of binary and ternary correlations are compared in figure~\ref{comp} where the finite limits at small distances are shown. The ternary order somewhat improves the binary quantum potential and leads to somewhat less binding behaviour.

This means that the Coulomb divergence at small distances is cured due to quantum fluctuations brought by many-body correlations. Please note that this is the opposite limit than the large-distance Coulomb behaviour which is cured by screening.  
This finite behaviour at small distance translates into a higher potential decay at large momenta. In fact, the Fourier transform of the binary potentials read
\ba
V_2(q)&={8\pi E_0 a_0\hbar^3\over q^3l } D\!\left (\!{q l\over 2 \hbar}\!\right )
={E_0 a_0}\!\left \{
\!\begin{array}{c}
{8\pi\hbar^4\over l^2 q^4}\!+\!o(q^{-6})\cr
{4\pi \hbar^2\over q^2}\!-\!{2\pi l^2\over 3} \!+\!o(q^2)
\end{array}
\right .
\label{V2ex}
\end{align}
and
\ba
V_3(q)&={32 E_0 a_0\hbar^3\over \sqrt{\pi}q^3 l} \int\limits_0^\infty du {\rm e}^{-u^2}D(u) \,{\rm ln}\left |{(u+{q l\over 2})\over (u-{q l\over 2}) }\right |
\nonumber\\
&={E_0 a_0}\left \{
\begin{array}{c}
{16\hbar^4\over l^2 q^4}+o(q^{-6})\cr
{4 \pi \hbar^2\over q^2}-{2 (2+\pi)\over 3} l^2+o(q^2)
\end{array}
\right .
\label{V3ex}
\end{align}
with the Dawson function $D(x)={\rm e}^{-x^2}\int\limits_0^x dy {\rm e}^{y^2}$.
Analogously one obtains for the Fermi potential
\ba
V_{2\rm f}(q)&={2\pi E_0 a_0 \hbar^2\over q^2} \left ( 
1+{p_{\rm F}^2-q^2\over 2 q p_{\rm F}}
{\rm ln}
\left |{p_{\rm F}+q\over pf-q}\right |
\right ) 
\nonumber\\
&={E_0 a_0}\left \{
\begin{array}{c}
{4\pi \hbar^4 \over 3 l_{\rm F}^2 q^4}+o(q^{-5})\cr
{4\pi \hbar^2\over q^2}-{4 \pi l_{\rm F}^2\over 3} +o(q^2)
\end{array}
\right .
\label{V2fex}
\end{align}
and
\ba
V_{3\rm f}(q)&={16\pi E_0 a_0\hbar^3 \over (4\!+\!\pi^2)q^3 l} \!\!\int\limits_0^1\!\!\! dx \!\left [x\!+\!\left (1 \!-\!x^2\right ) {\rm artanh} x\right ]{\rm ln}{|xp_{\rm F}\!+\!q|\over |xp_{\rm F}\!-\!q|}
\nonumber\\
&={E_0 a_0}\left \{
\begin{array}{c}
{16\pi \hbar^4\over (4+\pi^2)l_{\rm F}^2 q^4}+o(q^{-6})\cr
{4 \pi \hbar^2\over q^2}-{4 \pi (12+\pi^2)\over 3(4+\pi^2)} l_{\rm F}^2+o(q^2)
\end{array}
\right ..
\label{V3exf}
\end{align}

\begin{figure}
\includegraphics[width=8cm]{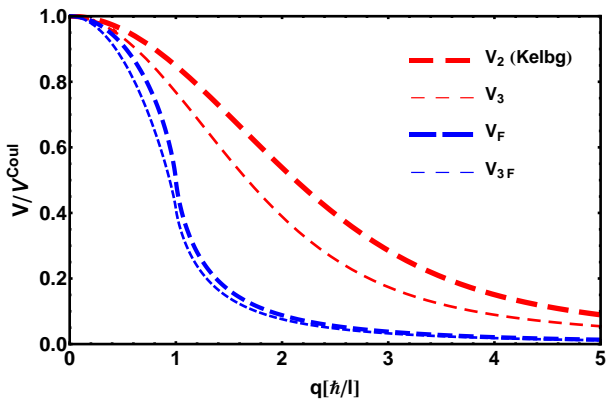}
\caption{
\label{compq} The ratio of the quantum potentials to the Coulomb one in momentum space. For the Kelbg and ternary potential the scale is $l=\hbar\sqrt{2\beta/m}$ and for the Fermi potential $l=l_{\rm F}=\hbar/p_{\rm F}$.
}
\end{figure}

These quantum potentials in momentum space are compared in figure~\ref{compq}. The potentials with Fermi correlations show a faster decay around the Fermi momentum $q=p_{\rm F}$ compared to the Maxwellian ones. We see that all the quantum potentials show a $V(q)\sim 1/q^4$ behaviour for large $q$ and a Coulomb behaviour at small $q$. Therefore these quantum potentials lead to $\alpha=4$ and according to the above discussions for the appearance of anomalies they vanish in all cases. Does this mean that the energy shift of the anomaly (\ref{deltaE}) does not exist? We will see how that will reappear quite ordinarily as the difference of the total energy calculated by quantum potentials compared with the one by the Coulomb potential. 

We consider the Hartree correlational energy corresponding to the lowest-order binary quantum potential for homogeneous systems 
\be
E_{\rm corr}=\frac 1 2 \int d^3 r' n(r-r') V(r')=\frac n 2 V(q=0)
\ee
which is the convolution of the potential with the particle density $n$.
According to (\ref{V2ex})-(\ref{V3exf}), the difference of one-particle energies between the quantum potentials and the Coulomb one is 
\be
E_1-E^c_1&=&\frac n 2 [V_{\rm quant}(q=0)-V^c(q=0)]
\nonumber\\
&=&-E_0 \left \{
\begin{array}{rlcc}
{1 \over 3\sqrt{\pi}}& {a_0\over l} &{\rm for}& V_2\cr
{2+\pi \over 3\pi^{3/2}}&{a_0\over l}  &{\rm for}& V_3\cr
{2\over 9\pi}&{a_0 \over l_{\rm F}}  &{\rm for}& V_{2\rm f}\cr
{2(12+\pi^2)\over 9\pi(4+\pi^2)}&{a_0 \over l_{\rm F}}  &{\rm for}& V_{3\rm f}
\end{array}
\right .
\label{e1}
\ee
where we have used the densities $n=1/\pi^{3/2}l^{3}$ for $V_{2,3}$ and $n=1/3\pi^2l_{\rm F}^3$ for $V_{2,3f}$. The kinetic energies subtract identically.

On the other hand we have the potential energy of a single particle $V(r=0)$ as indicated in figure~\ref{comp}. Adding the kinetic energies $3T/2=3 E_0 a_0^2/l^2$ for $V_{2,3}$ and $3 p_{\rm F}^2/10m=3 E_0 a_0^2/10 l_{\rm F}^2$ for $V_{2\rm f}$, we obtain the single-particle energy from (\ref{Fermi})-(\ref{ternf})
\be
E_1=E_0\left \{
\begin{array}{cc}
\sqrt{\pi} {a_0\over l}+3{ a_0^2\over l^2}  &{\rm for}\, V_{2}
\cr
\sqrt{8\over \pi}\,{\rm ln}(1+\sqrt{2}) {a_0\over l}+3{ a_0^2\over l^2}  &{\rm for}\, V_{3}
\cr
{\pi\over 4} {a_0\over l_{\rm F}}+{3\over 10}{a_0^2\over l_{\rm F}^2}  &{\rm for}\, V_{2\rm f}
\cr
{4\pi (1+2 {\rm ln}2)\over 3(4+\pi^2)} {a_0\over l_{\rm F}}+{3\over 10}{a_0^2\over l_{\rm F}^2}  &{\rm for}\, V_{3\rm f}
\end{array}
\right .
\label{e2}
\ee
Dividing now (\ref{e1}) by (\ref{e2}) and using the limit of $l\to 0$ corresponding to $\beta\to 0$ of chapter~\ref{eanom}, which is also the maximum ratio, we get
\be
{\langle E_1-E^c_1\rangle\over \langle E_1\rangle} =-{1\over c}
\ee
and therefore
\be
\Delta H={\langle E^c_1\rangle-\langle E_1\rangle}={\langle E^c_1\rangle\over 1+c}
\ee
with
\be
c=\left \{
\begin{array}{rccl}
{3\pi }&\approx& 9.24  &{\rm for}\, V_{2}
\cr
{6\pi\sqrt{2}\, {\rm ln}(1+\sqrt{2})\over (2+\pi)}&\approx& 4.57  &{\rm for}\, V_3
\cr
{9 \pi^2\over 8}&\approx& 11.10   &{\rm for}\, V_{2\rm f}
\cr
{6 \pi^2(1+2 {\rm ln}2)\over 12+\pi^2}&\approx& 6.46   &{\rm for}\, V_{3\rm f}
\end{array}
\right ..
\label{c}
\ee
Comparing with the exact ``anomalous'' result (\ref{deltaE}) we would expect $c=7$. The increasing quality of potentials from binary to ternary level is visible. The Maxwellian is less accurate than the Fermi correlation since we had considered Fermi ones in chapter II. We can now suggest to use this ration $c$ of the energy to the deviation of the energy with quantum potentials from the classical Coulomb one as a measure for the quality of the potential to represent quantum fluctuations.

\section{Summary}

The nonrelativistic quantum anomaly is investigated for combinations of momentum behaviour of potentials, dimensionality, and the order of perturbation. It is found that only for the energy an anomalous shift appears in three dimensions while in one dimension no anomaly is seen. In two dimensions the discussion can be performed analogously. It is seen that the quantum anomaly appears as the large momentum or short distance behaviour of the potential. Quantum potentials are proposed which describe quantum features on a classical level. These quantum potentials lead to a finite value at small distances and cure the Coulomb divergence. The consequence is that no quantum anomaly appears. In contrast, the deviation of the energy with quantum potentials from the energy with  the Coulomb potential reflects this anomalous energy shift. In this way the quantum anomalous behaviour is reformulated by normal quantum many-body correlations. It may be a hint that anomalies as such are a theoretical short cut to the right physics but can be formulated equivalently by a more refined many-body treatment. This of course needs further investigation on a more abstract level than considered here. The discussed quantum potentials might be usefull to describe the simulation of strongly correlated quantum systems by classical terms.

\appendix
\section{Integrals}
The occurring integrals in chapter II have the form
\ba
&I_n^{(D)}(\alpha)=\sum\limits_{p,k_1,...k_n} {\delta(k_1+...+k_n)\over k_1^\alpha...k_n^\alpha} {1\over 1+p^2}
\nonumber\\
&\times\left ({1\over 1\!+\!(\V p\!+\!\V k_1)^2}...{1\over 1\!+\!(\V p\!+\!\V k_1\!+...\!+\!\V k_n)^2}\!-\!{1\over (1\!+\!p^2)^n}\right )
\label{a1}
\end{align}
with $\sum_p=\int d^Dp/(2\pi)^D$
Introducing new variables $\V p_1=\V k_1+\V p$, $\V p_2=\V k_2+\V p_1$ etc. leads to
\ba
&I_n^{(D)}(\alpha)=\sum\limits_{p,p_1...p_{n-1}}
\left [{1\over 1+p_1^2}...{1\over 1+p_{n-1}^2}-{1\over (1+p^2)^{n-1}}\right ]
\nonumber\\
&
{1\over (\V p_1\!-\!\V p)^\alpha}{1\over (\V p_2\!-\!\V p_1)^\alpha}...{1\over (\V p_{n\!-\!1}\!-\!\V p_{n\!-\!2})^\alpha}{1\over (\V p\!-\!\V p_{n\!-\!1})^\alpha}{1\over (1\!+\!p^2)^2}
\label{in}
\end{align}

We are going to calculate the integrals for Coulomb potentials $\alpha=2$.

\subsection{3D}
Lets consider the integrals with increasing $n$ starting with the lowest non-vanishing one
\be
I_2^{(3)}(2)=\sum_{p,p_1}{1\over (\V p_1\!-\!\V p)^4} {1\over (1\!+\!p^2)^2}\left [{1\over 1\!+\!p_1^2}\!-\!{1\over 1\!+\!p^2}\right ]
\label{i2}
\ee
First performing the integrals about $p_1$
\be
&&{1\over 4 \pi^2}\int \limits_0^\infty d p_1 p_1^2\int\limits _{-1}^1 {d x\over (p_1^2+p^2-2 p_1 p x)^2}{1\over 1+p_1^2}
\nonumber\\
&&={1\over 4 \pi^2}\int \limits_{-\infty}^\infty d p_1 {p_1^2\over (p_1^2-p^2)^2}{1\over 1+p_1^2}
\label{i2a}
\ee
where for the second integral in (\ref{i2}) we do not have the last term in (\ref{i2a}). Using the residue calculus we circumvent the poles $p_1=\pm p$ by a semicircle with vanishing radius $\epsilon$ and obtain for (\ref{i2})
\ba
&I_2^{(3)}(2)
\nonumber\\
&={1\over 4 \pi^2}\sum\limits_p\left \{ 
{1\over (1\!+\!p^2)^2}\left [\frac 1 \epsilon {1\over 1\!+\!p^2}\!-\!{\pi\over (1\!+\!p^2)^2}\right ]\!-\!\frac 1 \epsilon {1\over (1\!+\!p^2)^3}\right \}
\nonumber\\
&=-{1\over 8 \pi^3}\int\limits_0^\infty dp {p^2\over (1+p^2)^4} =-{1\over 256 \pi^2}
\label{I322}
\end{align}
and the divergence cancels leading to a finite result.

All next order $n> 2$ integrals are divergent. This can be seen from the second part of (\ref{in}) which is a convolution and which can be written as Fourier transform of
\be
I_n^{(3)}(2)_{\rm right}=\int d^3 r \left ( {1\over 4 \pi r}\right )^n \sum \limits_p {1\over (1 +p^2)^{n+1}}. 
\ee
This is obviously divergent at small distance of the potential due to the powers $n>2$. The first part of (\ref{in}) instead is convergent as one can see, e.q. by calculating 
\ba
I_3^{(3)}(2)_{\rm left}&=\!\sum\limits_{p,p_1,p_2}{1\over (\V p_1\!-\!\V p)^2}{1\over (\V p_2\!-\!\V p_1)^2}{1\over (\V p\!-\!\V p_2)^2}{1\over (1\!+\!p^2)^2}
\nonumber\\&\times {1\over 1+p^2_1}{1\over 1+p^2_2}.
\end{align}
The integration over $p_1$ can be performed with the help of the Fourier transformation and shifting $\V k_1=\V p_1-\V p$
\ba
&\int {d^3p_1\over (2 \pi)^3}{1\over (\V p_1-\V p)^2}{1\over (\V p_2-\V p_1)^2}{1\over 1+p^2_1}
\nonumber\\
&=\int \!\! {d^3r d^3r'\over (4 \pi )^2}\!\!\!\int\limits_0^\infty \!\!\ {d k_1 \over 2 \pi^2 }{\sin{(k_1|r\!-\!r'|)}\over k_1} {\rm e}^{-i \V r\cdot (\V p_2\!-\!\V p)\!-\!i \V r'\cdot \V p\!-\!r'}.
\end{align}
Performing the integration over $k_1$ which gives $\pi/2$ and after shifting $ \V r=\V s +\frac 1 2 \V r'$ we can use
\be
\int\limits_0^\infty ds {s \sin{(s a)}\over s^2-{r'^2\over 4}}={\pi \over 2} \cos\left ({a r'\over 2}\right )
\ee
to obtain finally
\be
&&{1\over 4}\int \limits_0^\infty d r' {\cos {|p-p_2| r'\over 2}\over |p-p_2|}{\sin {|p+p_2| r'\over 2}\over |p+p_2|}{{\rm e}^{-r'}\over r'}
\nonumber\\
&&={\pi \over 8} {1\over p^2-p_2^2}
\ee
where we have used
\ba
&\int \limits_0^\infty d r' {\cos {(a r')}}{\sin {(b r')}}{{\rm e}^{-r'}\over r'}
\nonumber\\
&={\rm Im} \frac 1 4\int \limits_{-\infty}^\infty d r'{{\rm e}^{-|r'|}\over r'}  \left ({\rm e}^{i (a+b) r}+{\rm e}^{i (b-a) r} \right )
\nonumber\\
&=-\frac 1 4 {\rm Im}\, i\int\limits_\pi^0 d\phi (1+1)=\frac \pi 2.
\end{align}

\subsection{Integrals for $n=1$}
For $n=1$ the integral (\ref{a1}) takes the form
\ba
&I_1^{(D)}(\alpha)
=\sum\limits_{p,k_1} {\delta^D(\V k_1)\over k_1^\alpha} {1\over 1+p^2}
\left ({1\over 1\!+\!(\V p\!+\!\V k_1)^2}\!-\!{1\over (1\!+\!p^2)^n}\right )
\nonumber\\
&=-\sum\limits_{p,k_1} {\delta^D(\V k_1)\over k_1^\alpha} {k_1^2+2 \V p\cdot \V k_1 \over (1+p^2)^3}
=-\sum\limits_{p,k_1} {\delta^D(k_1)} {k_1^{2-\alpha}\over (1+p^2)^3}
\label{i1da}
\end{align}
where the integration over $p$ renders the scalar product zero.
A non-vanishing result is only for $\alpha=2$ which means that for one, two and three dimensions the case $D=\alpha-2$  is zero
\be
I_1^{(D)}(D-2)=0
\label{I1Da}
\ee
and for $D=\alpha$ the only finite result is
\be
I_1^{(2)}(2)=-{1\over 16 \pi^2}
\label{I122}
\ee
due to trivial integrations. In three dimensions there is no potential with $\alpha=3$.

\section{Quantum potentials\label{eff}}
In the following we indicate potentials without the units of $E_0 a_0$ by $\bar V$. A further possible fore-factor is added in the end corresponding to the demand that the Coulomb potential should be approached for large distances. 

\subsection{Binary potentials}
We calculate the convolution
\be
{\bar V}_2(r)=\int d^3 x \rho^{(2)}(x) {\bar V}^c(x) {\bar V}^c(\V x+\V r)
\label{aVc}
\ee
between the binary correlation $\rho^{(2)}$ and the Coulomb potential ${\bar V}^c(r)=1/r$.  The angular integration is trivial
\be
\!\int \!\!d\omega {\bar V}^c(\V x\!+\!\V r)\!=\!2\pi \!\!\!\int\limits_{-1}^1 \!\!{dz\over \sqrt{x^2\!+\!r^2\!+\!2 x r z}}\!=\!4\pi \!\!\left \{\begin{array}{c}
x^{-1}, x\!>\!r \cr
r^{-1}, x\!<\!r 
\end{array} \right .
\label{averV2}
\ee
such that
\be
{\bar V}_2(r)=4\pi \left ( \frac 1 r \int\limits_0^r d x x \rho^{(2)} (x) +\int\limits_r^\infty d x \rho^{(2)}(x)\right ).  
\label{aV2}
\ee

\subsubsection{Maxwellian correlations}
The Maxwellian correlation is the Fourier transform of the Maxwell distribution,
\be
\rho^{(2)}(r)=\int {d^3 p\over (2 \pi \hbar)^3}{\rm e}^{{i\over \hbar} \V r \V p-{p^2\over 2 m T}}={1\over \pi^{3/2} l^{3}} {\rm e}^{-x^2/l^2}
\ee
with the thermal wavelength $l^2=2 \hbar/mT=\lambda^2/\pi$.
One easily integrates (\ref{aV2}) with the result 
\ba
\bar V_2(r)&={2\over \sqrt{\pi} l r}
\left [
1-{\rm e}^{-\left ({r\over l}\right )^2}+\sqrt{\pi}\,{r\over l}\, {\rm erfc}
\left ({r\over l}\right ) 
\right ]
\nonumber\\
&={2\over \sqrt{\pi}  l}\left \{
\begin{array}{c}
{\sqrt{\pi}\over l}-{r\over l^2} +o(r^2)
\cr
{1\over r}+o(r^{-5})
\end{array}
\right .
\end{align}
which provides the fore-factor $E_0 a_0 \sqrt{\pi} l/2$ to obtain
(\ref{Kelbg}). This fore-factor is chosen such that the Coulomb potential appears for large $r$.

The Fourier transform into momentum space is in principle straight forward. However we will use the convolution structure since this will turn out to be helpful later for the ternary potentials. The potential (\ref{aVc}) translates into a product
\be
{\bar V}_2(q)=(\rho^{(2)} {\bar V}^c)(q) {\bar V}^c(-q)
\label{aV2q}
\ee
where we need to calculate the convolution $(\rho^{(2)} {\bar V}^c)(q)$. For Maxwellian correlations the angular integration is trivial and one gets 
\ba
 &
\int\!\!\! {d^3\!q'\over (2\pi \hbar)^3} {\bar V}^c(q')\rho^{(2)}(q\!-\!q')
=4\pi \hbar^2 \int \!\!\! {d^3q'\over (2\pi \hbar)^3}{{\rm e}^{-{l^2(q\!-\!q')^2\over 4\hbar^2}}\over q'^2}
\nonumber\\
&={2   \hbar \over \pi ql^2}\int\limits_0^\infty \!\!{d t\over t} \!\left ({\rm e}^{-\left ({q l\over 2\hbar} \!-\!t\right )^2}\!-\!{\rm e}^{-\left ({q l\over 2\hbar} \!+\!t\right )^2}\right )
={4\hbar \over \sqrt{\pi} l^2 q} D\left ({q l\over 2\hbar}\right )
\label{rV}
\end{align}
with the Dawson integral $D(x)={\rm e}^{-x^2}\int\limits_0^x dy {\rm e}^{y^2}$. With (\ref{aV2q}) we obtain
\ba
{\bar V}_2(q)={2\over \sqrt{\pi} l} {8 \pi \hbar^3 \over q^3 l^2} D\!\left (\!{q l\over 2\hbar}\!\right )={2\over \sqrt{\pi} l} {4\pi \hbar^2\over q^2} \!\left \{\!
\begin{array}{c}
1\!-\!{l^2\over 6 \hbar^2} q^2\!+\!o(q^3)\cr
{2\hbar^2\over l^2 q^2} \!+\!o(q^{-3})
\end{array}
\right .
\end{align}
and we see again the fore-factor 
$E_0 a_0 \sqrt{\pi} l/2$ to get the expression (\ref{V2ex}) including the expansions.

\subsubsection{Fermi correlations}
With Fermi correlations as Fourier transform of the Fermi function,
\ba
\rho_{2\rm f}(x)=\int\limits_{p\le p_{\rm F}}\!\!\!\! {d^3 p\over (2 \pi \hbar)^3}{\rm e}^{{i\over \hbar} r p}={1 \over 2 \pi^2 r^3} \left (\sin{r\over l_{\rm F}}\!-\!{r\over l_{\rm F}} \cos{r\over l_{\rm F}}\right ),
\end{align}
with $l_{\rm F}=\hbar/p_{\rm F}$,
one obtains for (\ref{aV2})
\ba
V_{2f}(r)&={1 \over \pi r l_{\rm F}}
\left \{
2-{\cos{\bar r}}- {\sin{\bar r}\over \bar r} +{\bar r} \left [\frac \pi 2-Si({\bar r}\right ]
\right \}
\nonumber\\
&=\left \{
\begin{array}{c}
{1\over 2 l_{\rm F}^2}-{r\over 3\pi l_{\rm F}^3} +o(r^2)
\cr
{2\over \pi r l_{\rm F}}+o(r^{-3})
\end{array}
\right .
\end{align}
with $\bar r=r/l_{\rm F}$. We see that the fore-factor to be chosen is here $E_0 a_0 \pi l_{\rm F}/2$ to obtain (\ref{Fermi}).

The Fourier transform we calculate analogously to the Maxwellian with
\ba
(\rho_{2\rm f} {\bar V}^c)(p)&=\int {d^3q\over (2\pi \hbar)^3} {\bar V}^c(p-q)\rho_{2f}(q)
\nonumber\\
&={1\over 2 \pi \hbar p}\int\limits_0^{p_{\rm F}} dq q\,{\rm ln}{(p\!+\!q)^2\over (p\!-\!q)^2} 
\nonumber\\
&={p_{\rm F}\over 2 \pi \hbar} \left [ 2+\left ( {p_{\rm F} \over p}-{p\over p_{\rm F}}\right ){\rm ln}{|p+p_{\rm F}|\over |p-p_{\rm F}|}\right ]
\label{rVf}
\end{align}
where we have used the trivial angular integration
\ba
\int d\Omega {\bar V}^c(p-q)=\int\limits_{-1}^1 \!\!dx {8\pi^2\hbar^2\over q^2\!+\!p^2\!-\!2 p q x}={4\pi^2\hbar^2\over p q}{\rm ln}{(p\!+\!q)^2\over (p\!-\!q)^2}. 
\label{ang}
\end{align}
With the help of (\ref{rVf}) we obtain for (\ref{aV2q})  
\ba
{\bar V}_{2f}(q)&={2 \hbar^2\over l_{\rm F} q^2} 
\left [ 2+\left ( {p_{\rm F} \over p}-{p\over p_{\rm F}}\right )
{\rm ln}{|p+p_{\rm F}|\over |p-p_{\rm F}|}\right ]
\nonumber\\
&={8\hbar^2\over l_{\rm F} q^2} \!\left \{\!
\begin{array}{c}
1\!-\!{q^2\over 3 p_{\rm F}^2} \!+\!o(q^4)
\cr
{p_{\rm F}^2\over 3 q^2} \!+\!o(q^{-4})
\end{array}
\right .
\end{align}
which provides again the fore-factor $E_0 a_0 \pi l_{\rm F}/2$ as above to get finally
(\ref{V2fex}).

\subsection{Ternary potentials}

The convolution structure of the ternary potentials (\ref{ternary}) 
suggests to calculate them in momentum space
\ba
{\bar V}_3(q)=V(-q) \int{d^3p\over (2\pi\hbar)^3}(\rho^{(2)} {\bar V}^c)(p) \rho^{(2)}(p) {\bar V}^c(q-p)
\label{tern1}
\end{align}
where we can conveniently use the results of the foregoing chapter (\ref{rV}) or (\ref{rVf}), respectively. 

\subsubsection{Maxwellian correlations}

Introducing the simple angular integration (\ref{ang}) into (\ref{tern1}) and
using (\ref{rV}) we obtain 
\ba
{\bar V}_{3}(q)&={2\over \sqrt{\pi}}\int \limits_0^\infty du {\rm e}^{-u^2}D(u)
{\rm ln}{(u+{q l\over 2 \hbar})^2\over (u-{q l\over 2 \hbar})^2}
\nonumber\\
&={4\pi \hbar^2\over l^2 q^2} \!\left \{\!
\begin{array}{c}
1\!-\!{(2+\pi)q^2l^2 \over 6\pi \hbar^2} \!+\!o(q^4)
\cr
{4\hbar^4\over \pi  q^2 l^4} \!+\!o(q^{-4})
\end{array}
\right .
\end{align}
where we used the integrals
\ba
\int\limits_0^\infty {\rm e}^{-u^2} D(u)
\begin{pmatrix}
u\cr u^{-1}\cr u^{-3}
\end{pmatrix}
=
{\sqrt{\pi}\over 8}
\begin{pmatrix}
1\cr \pi \cr -2 (2+\pi)
\end{pmatrix}
\label{ints}
\end{align}
for the small and large-$q$ expansions.
The comparison with the Coulomb potential for small $q$ provides the fore-factor $E_0 a_0 l^2$ to get just (\ref{V3ex}). 

For the potential in spatial domain we integrate directly (\ref{ternary}) with a trivial renaming $\V x_1\to \V x, \V x_2\to \V y_2-\V x$
\ba
 V_3(r)\!=\!\!\!\int\!\! d^3 x d^3y \rho^{(2)}(x_1)V^c(x_1)V^c(y_2) \rho^{(2)} (\V y\!-\!\V x) V^c(\V y\!+\!\V r).
\end{align}
The angular integration of $y$ is given by (\ref{averV2}) and the one of $x$ by
\be
\int d\Omega_x \rho^{(2)}(\V y-\V x)={2 {\rm sinh}{2 x y\over l^2}\over \pi^2 l^4 x y} {\rm e}^{-(x^2+y^2)\over l^2}.
\ee
The $|x|$ integration yields error functions. Using $y=l t$ the remaining integration reads
\be
V_{3}(r)={2^{3/2}\over \sqrt{\pi} l^3}\! \left (\,\int\limits_{r/l}^\infty \!{d t\over t}\!+\!{l\over r} \!\int\limits_0^{r/l}\! d t\right ) {\rm e}^{-{t^2\over 2}} {\rm erf} \left ({t\over \sqrt{2}}\right ).
\ee 
The second integral can be made analytically observing that one can perform a partial integration $\bar r =r/l$
\ba
&\int\limits_0^{\bar r}\! d t 
\left [ 
\sqrt{\pi \over 2}  {\rm erf} 
\left ({t\over \sqrt{2}}\right )
\right ]' 
{\rm erf} \left ({t\over \sqrt{2}}\right )
=
\left . \sqrt{\pi \over 2}  {\rm erf}^2 \left ({t\over \sqrt{2}}\right )
\right |_0^{\bar r}
\nonumber\\
&-\sqrt{\pi \over 2}  \int\limits_0^{\bar r} \!\!d t\, {\rm erf} 
\left (\!{t\over \sqrt{2}}\!\right ) 
\sqrt{2\over \pi} {\rm e}^{-{t^2\over 2}}
=\sqrt{\pi\over 8}  {\rm erf}^2 \left (\!{{\bar r}\over \sqrt{2}}\!\right ).
\end{align}
The last identity appears just observing that the second term from the partial integration is just the negative of the desired integral itself. Finally we obtain
\ba
{\bar V}_3(r)&={1\over l^3}
\left [
{2^{3/2} r\over \sqrt{\pi} l}\int\limits_{r/l}^\infty \!{d t\over t} {\rm e}^{-{t^2\over 2}} {\rm erf} \left ({t\over \sqrt{2}}\right )
+{\rm erf}^2 \left (\!{r\over \sqrt{2}l}\!\right )
\right]
\nonumber\\
&
={1\over l^2}\left \{\begin{array}{c}
{\sqrt{8}{\rm ln}(1+\sqrt{2})\over\sqrt{\pi} l}-{2\over \pi} {r\over l^2}+{2r^2\over 3 \pi l^3}+o(r^3)
\cr
{1\over r} +o(r^{-2})
\end{array}
\right ..
\end{align}
This shows again the fore-factor $E_0 a_0 l^2$ to get (\ref{tern}).

\subsubsection{Fermi correlations}
For Fermi correlations we introduce (\ref{rVf}) and (\ref{ang}) into (\ref{tern1}) to obtain 
\ba
&{\bar V}_{3f}(q)={4p_{\rm F}^3\over \pi q^3}\!\!\int\limits_0^1\!\! d x [x\!+\!(1\!-\!x^2)\, {\rm artanh}(x)]\,{\rm ln}{|xp_{\rm F}\!+\!q|\over |xp_{\rm F}\!-\!q| }
\nonumber\\
&={p_{\rm F}^2(4+\pi^2)\over 4\pi^2\hbar^2}
\left \{
\begin{array}{c}
{16\hbar^4\over (4+\pi^2)l_{\rm F}^2 q^4}+o(q^{-6})\cr
{4\pi \hbar^2\over q^2}\!-\!{4\pi \hbar^2\over 3 p_{\rm F}^2}{(12+ \pi^2)\over (4+\pi^2)}+o(q^2)
\end{array}
\right ..
\label{V3f1}
\end{align}
Choosing the fore-factor as $E_0e_0 4\pi^2\hbar^2/p_{\rm F}^2(4+\pi^2)$ such that for small momentum $q$ the Coulomb result appears corresponding to large distance behaviour, we obtain just (\ref{V3exf}).

The first term of the small $q$-expansion of (\ref{V3f1}) and and large $q$-expansion can be performed directly using
\ba
\int\limits_0^1\!\!\! dx \!\left [x\!+\!\left (1 \!-\!x^2\right ) {\rm artanh} (x)\right ]
\begin{pmatrix}
x \cr 1 \cr x^{-1}
\end{pmatrix}
=
\begin{pmatrix}
\frac 1 2 \cr \frac 1 3 (1+2{\rm ln}2) \cr \frac 18 (4+\pi^2)
\end{pmatrix}.
\label{x}
\end{align}

The second term in the small-$q$ expansion of (\ref{V3exf}) deserves special attention. The needed integral with $x^{-3}$ in (\ref{x}) would diverge. The reason is a tricky order of principal value integrations. The best way to solve this problem is to consider a Debye potential $\exp{(-\kappa r/\hbar)}/r$ with a vanishing $\kappa$ in the angular integration (\ref{ang}) instead of the Coulomb potential $\sim 1/r$,
\ba
\int d\Omega {\bar V}^c(\V q-\V p)&=&\int\limits_{-1}^1 \!\!dx {8\pi^2\hbar^2\over q^2\!+\!p^2\!-\!2 p q x+\kappa^2}
\nonumber\\
&=&{4\pi^2\hbar^2\over p q}{\rm ln}{(p\!+\!q)^2+\kappa^2\over (p\!-\!q)^2+\kappa^2}.
\label{ange}
\end{align}
This leads instead of (\ref{rVf}) to 
\ba
(\rho_{2\rm f} {\bar V}^c)(p)
=&{p_{\rm F}\over 2 \pi \hbar} \left [ 2
-{2 e\over x} {\rm arctan}{2e\over e^2+x^2-1}
\right .
\nonumber\\
&\left . 
+\left ( {e^2-x^2+1\over 2 x}\right ){\rm ln}{(1+x)^2+e^2\over (1-x)^2+e^2}
\right ]
\label{rVfe}
\end{align}
with $x=p/p_{\rm F}$ and $e=\kappa/p_{\rm F}$. Though the limit $e\to 0$ gives (\ref{rVf}) the $x$-integral in (\ref{V3f1}) diverges if performed after this limit and is finite when the limit is performed after the integration. To see this, we consider the large-$q$ expansion
\be
{1\over q^3}{\rm ln}{|xp_{\rm F}\!+\!q|\over |xp_{\rm F}\!-\!q| }={2\over x q^2 p_{\rm F}}+{2 \over 3 x^3p_{\rm F}^3}+...
\ee
and have with (\ref{rVfe}) instead of (\ref{V3f1})
\ba
&{\bar V}_{3f}(q)={8\over \pi}\!\!\int\limits_0^1\!\! d x 
\left [
x
-{e}\, {\rm arctan}{2e\over e^2+x^2-1}
\right .
\nonumber\\
&\left .
+\left ( {e^2\!-\!x^2\!+\!1\over 4}\right ){\rm ln}{(1\!+\!x)^2\!+\!e^2\over (1\!-\!x)^2\!+\!e^2}
\right ] \left ({p_{\rm F}^2\over x q^2 }\!+\!{1 \over 3 x^3}\!+\!o(q^2)\right ). 
\label{V3f2}
\end{align}
The first term $\sim 1/x$ is convergent in the $e\to 0$ limit according to (\ref{x}).
For the second problematic $q^0$ term $\sim 1/x^3$ we first integrate and then perform the limit with the result
\ba
{\bar V}_{3f}(q)={p_{\rm F}^2\over q^2}\left ({4\over \pi}\!+\!\pi\right )\!-\!\left ({4\over \pi}\!+\!{\pi\over 3}\right )\!+\!{8\over 3 \pi} e \!+\!o(e^2,q^2)
\end{align}
which after $e\to 0$ gives the expansion (\ref{V3f1}) and (\ref{V3exf}).

The form (\ref{V3f1}) is convenient for the Fourier transform which yields
\ba
&{\bar V}_{3f}(r)=\int {d^3 q\over (2\pi \hbar)^3} {\bar V}_{3f}(q)
\nonumber\\
&={p_{\rm F}^2\over \pi^3\hbar^2 r}\!\!\int\limits_0^\infty \!\!d y {\sin{y \bar r}\over y^2}\!\!\int\limits_0^1 \!\!dx
[x\!+\!(1\!-\!x^2)\, {\rm artanh}(x)]\,{\rm ln}{(x\!+\!y)^2\over (x\!-\!y)^2}
\end{align}
with $y=q/p_{\rm F}$ and $\bar r=r p_{\rm F}/\hbar$. The $y$-integration can be performed
\ba
\int\limits_0^\infty \!\!d y {\sin{y \bar r}\over y^2}\,{\rm ln}{(x\!+\!y)^2\over (x\!-\!y)^2}&={\pi\over x}\left [2\!+\!\pi x \bar r\!-\!2 \cos{x\bar r}\!-\!2 x\bar r Si(x\bar r)\right ]
\nonumber\\
&=\left \{\begin{array}{c}
\pi^2 \bar r-\pi x \bar r^2+o(\bar r^3)\cr
{2\pi \over x}+o(\bar r^{-1})
\end{array}
\right .
\end{align}
with the sinus integral  $Si(x)=\int_0^xdt \sin{t}/t$.
Using (\ref{x}) the needed fore-factor can be seen from the $r\to 0$ expansion
\ba
{\bar V}_{3f}(r)&=
{2 p_{\rm F}^2\over \pi^2\hbar^2 r}\!\!\int\limits_0^1 \!\!dx
[x\!+\!(1\!-\!x^2)\, {\rm artanh}(x)]\,\left [ x^{-1}+o(r)\right ]
\nonumber\\
&
={p_{\rm F}^2(4+\pi^2)\over 4\pi^2\hbar^2}
\left \{
\begin{array}{c}
{4 \pi (1+2{\rm ln} 2) \over 3 (4+\pi^2)l_{\rm F}}-{2 r\over (4+\pi^2) l_{\rm F}^2}+o(r^3)
\cr
{1\over r} +o(r^{-2})
\end{array}
\right .,
\end{align}
again to be $E_0e_0 4\pi^2\hbar^2/p_{\rm F}^2(4+\pi^2)$ in order to obtain the Coulomb potential for large distances which all together provides (\ref{ternf}).

\bibliography{entropy,bose,kmsr,kmsr1,kmsr2,kmsr3,kmsr4,kmsr5,kmsr6,kmsr7,delay2,spin,spin1,refer,delay3,gdr,chaos,sem3,sem1,sem2,short,cauchy,genn,paradox,deform,shuttling,blase,spinhall,spincurrent,tdgl,pattern,zitter,graphene,quench,msc_nodouble,iso,march,weyl,anomal}
\bibliographystyle{../../script/apsrev_meins}

\end{document}